\pdfoutput=1
\documentclass[11pt]{article}

\usepackage[margin=1in]{geometry}
\usepackage{amsmath,amssymb,amsfonts}
\usepackage{graphicx}
\usepackage{booktabs}
\usepackage{siunitx}
\usepackage{caption}
\usepackage{subcaption}
\usepackage[hidelinks]{hyperref}
\usepackage{xcolor}
\usepackage{authblk}
\usepackage{microtype}

\graphicspath{{figures/}{figures/test1/}{figures/test2/}}

\newcommand{\uvec}{\mathbf{u}}
\newcommand{\eps}{\varepsilon}

\newcommand{\ddx}[1]{\partial_x #1}
\newcommand{\ddy}[1]{\partial_y #1}
\newcommand{\ddt}[1]{\partial_t #1}
\newcommand{\lap}{\nabla^2}
\newcommand{\relL}{\text{rel-}L_2}

\hypersetup{
  pdftitle={An Artificial-Compressibility Physics-Informed Neural Network for the Unsteady Incompressible Navier--Stokes Equations},
  pdfauthor={Aytekin Bayram \c{C}\i b\i k},
  pdfsubject={Physics-informed neural networks; artificial compressibility; incompressible Navier--Stokes; data assimilation},
  pdfkeywords={PINN, artificial compressibility, Navier-Stokes, data assimilation, inverse problems}
}

\title{An Artificial-Compressibility Physics-Informed Neural Network\\
for the Unsteady Incompressible Navier--Stokes Equations}
\author{Aytekin Bayram \c{C}\i b\i k}
\affil{Department of Mathematics, Gazi University, Ankara, T\"urkiye}
\date{\today}

\begin{document}
\maketitle

\begin{abstract}
We study a physics-informed neural network (PINN) for the unsteady, two-dimensional
incompressible Navier--Stokes equations in which the stiff divergence-free
constraint is replaced by an artificial-compressibility (AC) relaxation governed by
a single scalar parameter $\eps$. The relaxation reintroduces a pressure time
derivative, converting a differential-algebraic constraint into an ordinary
residual that a PINN can minimise directly. On the Taylor--Green vortex, which
admits a closed-form unsteady solution, we quantify the effect of $\eps$: the
residual divergence scales as $\eps\,|\partial_t p|$, so larger $\eps$ raises both
the divergence and the velocity error, and both decrease monotonically and
saturate as $\eps$ is reduced. Averaged over random seeds, well-converged
fixed-$\eps$ training is stable across five decades of $\eps$ and already reaches
sub-$0.5\%$ velocity error on this smooth benchmark; an $\eps$-continuation
schedule with adaptive weighting improves this by a further factor of a few but is
not strictly necessary here. On the lid-driven cavity at $Re=100$ the method
recovers the Ghia~et~al. centerline profiles to $7.7\%$ relative-$L_2$ once the
physical time window is long enough for the primary vortex to reach steady state.
On the $Re=100$ cylinder wake the plain forward AC-PINN collapses to the steady
symmetric branch and does not reproduce von K\'arm\'an shedding; assimilating a few
hundred sparse velocity sensors from a boundary-layer-resolved finite-element
reference (whose Strouhal number, $0.176$, we bring close to the $0.164$--$0.172$
literature band by resolving the separating shear layer, though it remains just
above it) recovers the unsteady vortex street to $7\%$ over the wake and its
shedding frequency to within $3\%$ of that same reference --- a bound set by the
reference's own fidelity rather than an independent validation against the true
flow. Jointly inferring the
viscosity is harder: with wake sensors alone $\nu$ drifts to nearly an order of
magnitude too small, but placing sensors in the boundary layer---where the viscous
term lives---makes it identifiable, improving the estimate fivefold. In keeping with
an honest methods narrative, we report which techniques --- variable scaling,
adaptive weighting, $\eps$-continuation, boundary-layer-adaptive sampling, and data
assimilation --- prove necessary, and where they fall short.
\end{abstract}

\section{Introduction}
\label{sec:intro}
Physics-informed neural networks (PINNs)~\cite{raissi2019} minimise the residual of
a governing PDE, evaluated by automatic differentiation at collocation points, over
the parameters of a neural network, and have been applied across an increasingly
broad range of forward and inverse problems in fluid
mechanics~\cite{jin2021nsfnets,karniadakis2021}. For the incompressible Navier--Stokes equations
the pressure is not an evolution variable: it is fixed instantaneously and
non-locally by the divergence-free constraint. A PINN must therefore satisfy this
constraint as a global differential-algebraic condition, which couples every
collocation point and is a recognised source of slow or stalled
training~\cite{wang2021,krishnapriyan2021}. Several
remedies exist---adding a penalty on the divergence, parameterising the velocity
through a stream function so incompressibility holds by construction, projecting
onto a divergence-free space, or supplying additional interior data to anchor the
solution at high Reynolds number~\cite{jin2021nsfnets,hu2025}---each trading
accuracy, generality, or cost.

A second, largely orthogonal line of work has targeted not the constraint itself
but the pathologies of PINN training more broadly: gradient-based adaptive loss
weighting to balance stiff and non-stiff terms~\cite{wang2021,xiang2021}, respecting
the temporal causal structure of evolution equations~\cite{wang2024causal},
variable-scaling and non-dimensionalisation strategies that improve the
conditioning of the underlying neural tangent
kernel~\cite{ko2025vspinn,wang2022ntk}, and architectural changes such as
Fourier-feature embeddings or residual-adaptive networks that mitigate spectral
bias~\cite{tancik2020,wang2024piratenets}. A consolidated account of this toolbox is
given in~\cite{wang2023guide}. These techniques are complementary to, rather than a
substitute for, resolving the incompressibility constraint itself.

We revisit instead the \emph{artificial-compressibility} idea~\cite{chorin1967,chorin1968},
long standard in classical CFD, in which the continuity equation is relaxed by a
pressure time derivative scaled by a small parameter $\eps$. A few recent PINN
studies use a compressible or artificial-viscosity formulation toward a different
end, e.g.\ high-Reynolds-number turbulence closure or genuinely compressible
flow~\cite{song2026}, but the accuracy--trainability trade-off of $\eps$ itself has
not, to our knowledge, been mapped out on canonical incompressible benchmarks. In
the PINN setting
this has an appealing consequence: every governing equation acquires a time
derivative, so the residual system becomes purely local and no global algebraic
constraint remains, at the price of an $\eps$-controlled modelling error. Our aim is
not to claim state-of-the-art accuracy but to characterise this trade-off honestly
and to map where lightweight training aids suffice and where they do not.

We make three contributions. First, on the Taylor--Green vortex---which admits a
closed-form unsteady solution---we quantify the role of $\eps$ over five decades and
multiple seeds, showing the divergence scales as $\eps\,|\partial_t p|$ and that
well-converged fixed-$\eps$ training is stable and sub-$0.5\%$ accurate without
elaborate fixes. Second, on the $Re=100$ lid-driven cavity we recover the
Ghia~et~al.~\cite{ghia1982} centerline profiles to $7.7\%$, and identify the
physical time horizon as the controlling factor. Third, on the $Re=100$ cylinder
wake we show that a data-free forward AC-PINN collapses to the steady symmetric
branch and does not reproduce vortex shedding---a collapse also reported for
related constrained PINN architectures on this benchmark~\cite{hu2025}---delimiting
the method's reach and
motivating causal training and data assimilation. Throughout we report, in the
spirit of a methods paper, exactly which of a graded toolbox of techniques each
problem required.

\section{Governing Equations}
\label{sec:governing}
We consider the velocity field $\uvec = (u,v)$ and pressure $p$ satisfying the
two-dimensional unsteady incompressible Navier--Stokes equations on a domain
$\Omega \times (0,T]$,
\begin{align}
\ddt{u} + u\,\ddx{u} + v\,\ddy{u} + \ddx{p} - \nu\lap u &= 0, \label{eq:momx}\\
\ddt{v} + u\,\ddx{v} + v\,\ddy{v} + \ddy{p} - \nu\lap v &= 0, \label{eq:momy}\\
\ddx{u} + \ddy{v} &= 0, \label{eq:cont}
\end{align}
with kinematic viscosity $\nu$. The continuity constraint~\eqref{eq:cont} contains
no time derivative of pressure; the pressure is instead determined
instantaneously and non-locally by the incompressibility condition. For a PINN
this manifests as a stiff differential-algebraic constraint that couples all
collocation points through a global divergence-free requirement and is a
well-documented source of slow or stalled training~\cite{wang2021,krishnapriyan2021}.

\section{Artificial-Compressibility Formulation}
\label{sec:ac}
Following Chorin's artificial-compressibility method~\cite{chorin1967,chorin1968}, we replace
the constraint~\eqref{eq:cont} by the relaxed continuity equation
\begin{equation}
\eps\,\ddt{p} + \bigl(\ddx{u} + \ddy{v}\bigr) = 0,
\qquad \eps > 0. \label{eq:ac}
\end{equation}
Equation~\eqref{eq:ac} couples the pressure to the local dilatation through the
single scalar $\eps$, which plays the role of an inverse squared artificial sound
speed, exactly as in Chorin's original pseudo-compressibility iteration and its
many descendants in classical CFD~\cite{shen1994,cappanera2025}. In the limit
$\eps \to 0$ the incompressible constraint~\eqref{eq:cont} is
recovered exactly --- a convergence to the incompressible (Leray) weak solution
that has been made rigorous in the applied-analysis
literature~\cite{temam1969,donatelli2008}; for finite $\eps$ the pressure
equilibrates the divergence on a
time scale $\mathcal{O}(\eps)$, admitting a residual divergence of the same order.

The benefit for PINN training is that~\eqref{eq:ac} is a genuine evolution
equation: every governing equation now carries a time derivative, so the residual
system is local in the same sense as the momentum equations and no global
algebraic constraint remains. The cost is a modelling error controlled by $\eps$.
Quantifying this trade-off --- accuracy versus optimisation stiffness --- is the
central object of Test~1 (Section~\ref{sec:test1}).

\section{Network, Scaling, and Loss}
\label{sec:network}
A fully-connected $\tanh$ network $\mathcal{N}_\theta:(x,y,t)\mapsto(u,v,p)$
represents the fields, with an optional random Fourier-feature input
embedding~\cite{tancik2020} for
multi-scale flows. Inputs and outputs are affinely rescaled to $[-1,1]$
(VS-PINN-style variable scaling~\cite{ko2025vspinn}), so the network operates in a
well-conditioned regime while PDE residuals are evaluated on the de-scaled
physical fields via automatic differentiation, which materially improves the
conditioning of the neural tangent kernel that governs PINN training
dynamics~\cite{wang2022ntk,ko2025vspinn}.

The composite objective is
\begin{equation}
\mathcal{L}(\theta) = w_{\mathrm{pde}}\,\mathcal{L}_{\mathrm{pde}}
                    + w_{\mathrm{ic}}\,\mathcal{L}_{\mathrm{ic}}
                    + w_{\mathrm{bc}}\,\mathcal{L}_{\mathrm{bc}},
\label{eq:loss}
\end{equation}
where $\mathcal{L}_{\mathrm{pde}}$ is the mean-squared residual of
\eqref{eq:momx}--\eqref{eq:momy} and~\eqref{eq:ac} at interior collocation points,
and $\mathcal{L}_{\mathrm{ic}}$, $\mathcal{L}_{\mathrm{bc}}$ are mean-squared
data-fit terms enforcing the initial and boundary conditions. Boundary supervision
is applied per component, so problems with no known wall pressure supervise the
velocity only.

\subsection{The progressive toolbox}
\label{sec:toolbox}
We begin from the plain AC-PINN and introduce each of the following techniques only
when a test fails without it, so that Section-by-section we can state precisely
which were necessary (see~\cite{wang2023guide} for a consolidated account of this
toolbox more broadly):
\begin{enumerate}
  \item \textbf{Variable scaling} --- always enabled; inexpensive and uniformly
        beneficial.
  \item \textbf{Adaptive loss weighting} --- gradient-norm balancing of the
        IC/BC terms against the PDE term~\cite{wang2021}, refreshed periodically.
  \item \textbf{$\eps$-continuation} --- training proceeds through a geometrically
        decreasing ladder of $\eps$ values, warm-starting each stage from the last,
        to reach small $\eps$ without the stiffness of starting there.
  \item \textbf{Causal pseudo-time weighting} --- for strongly time-dependent
        flows, weighting residuals so earlier times are fit before later ones
        \cite{wang2024causal}.
  \item \textbf{Augmented-Lagrangian enforcement} --- promoting the divergence
        constraint from a soft penalty to a Lagrangian term when accuracy plateaus.
\end{enumerate}
Optimisation uses Adam followed by an optional L-BFGS polish; collocation points
are sampled uniformly in space--time with rejection sampling around embedded
geometry. All experiments run on a single GPU.

\section{Test 1: Taylor--Green Vortex}
\label{sec:test1}
\subsection{Problem}
The Taylor--Green vortex is an exact unsteady solution of
\eqref{eq:momx}--\eqref{eq:cont} on $[0,2\pi]^2$,
\begin{equation}
u = \cos x \sin y\, F(t), \quad
v = -\sin x \cos y\, F(t), \quad
p = -\tfrac14(\cos 2x + \cos 2y)\, F(t)^2,
\label{eq:tg}
\end{equation}
with $F(t) = e^{-2\nu t}$. Because~\eqref{eq:tg} is known in closed form for all
time, it yields clean quantitative error metrics and isolates the effect of $\eps$.
We take $\nu = 0.05$ over $t \in [0,1]$; initial and boundary conditions are
sampled from the exact fields (all three components supervised, the exact pressure
being available).

\subsection{Training configuration}
The network has $5$ hidden layers of width $128$ ($66{,}947$ parameters). It is
trained with $\eps$-continuation over the geometric ladder
$\eps \in \{0.1,\,2.15\times10^{-2},\,4.64\times10^{-3},\,10^{-3}\}$, with
$4000$ Adam epochs per stage ($16{,}000$ total, warm-started between stages)
followed by an L-BFGS polish, using $6000$ interior and $1500$ each of
initial/boundary collocation points resampled per stage. Gradient-norm adaptive
weighting balanced the IC/BC terms against the PDE residual, converging to
weights $w_{\mathrm{ic}}\!\approx\!2.1$, $w_{\mathrm{bc}}\!\approx\!2.4$. Total
training time was $1199$\,s on a single NVIDIA GPU (CUDA, \texttt{torch~2.4}).

\subsection{Accuracy}
\begin{figure}[t]
  \centering
  \includegraphics[width=0.92\textwidth]{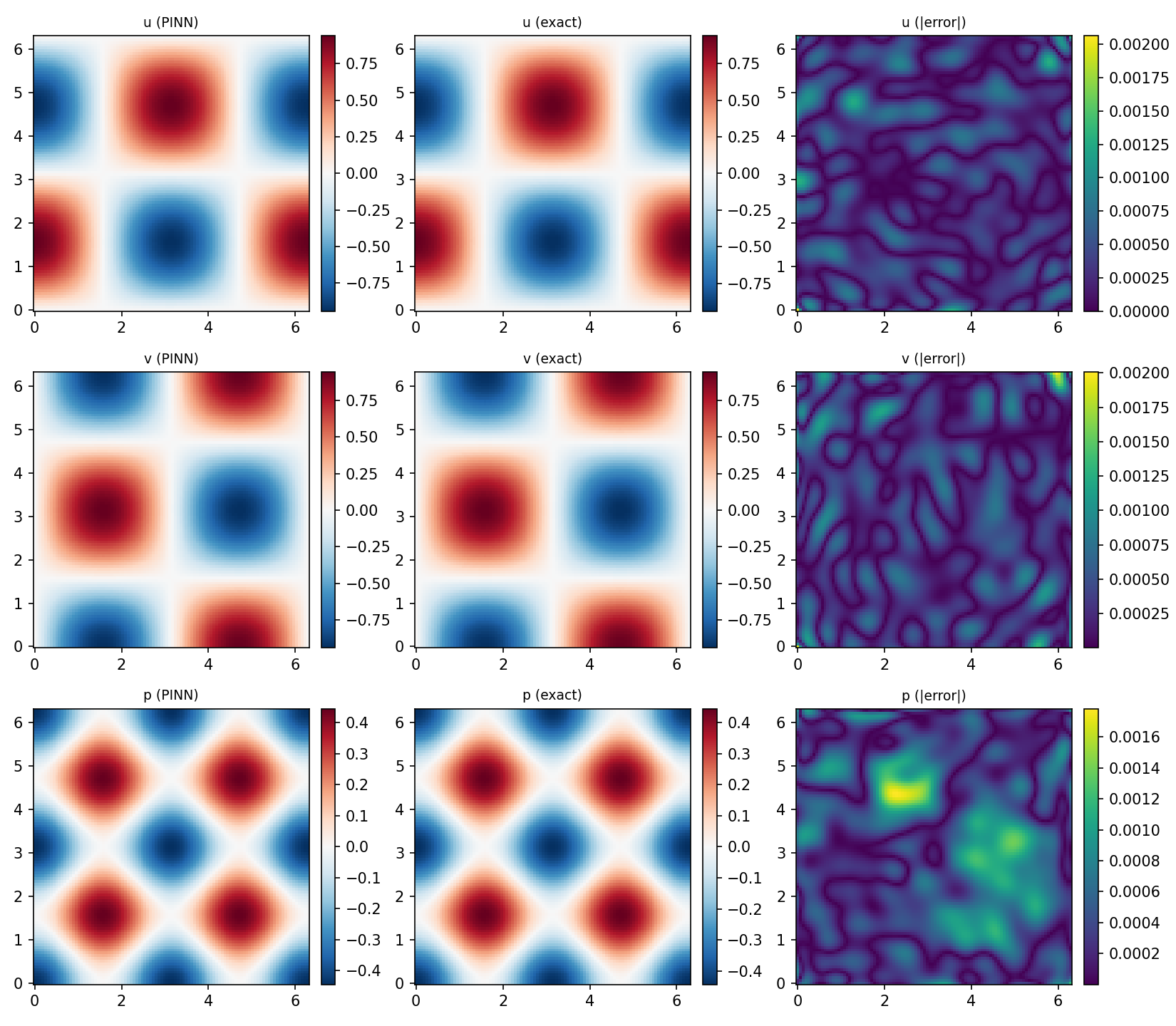}
  \caption{Taylor--Green vortex at $t=0.6$: AC-PINN prediction (left), exact
    solution~\eqref{eq:tg} (centre), and pointwise absolute error (right) for
    $u$, $v$, and $p$.}
  \label{fig:tg-fields}
\end{figure}

\begin{figure}[t]
  \centering
  \begin{subfigure}{0.48\textwidth}
    \includegraphics[width=\textwidth]{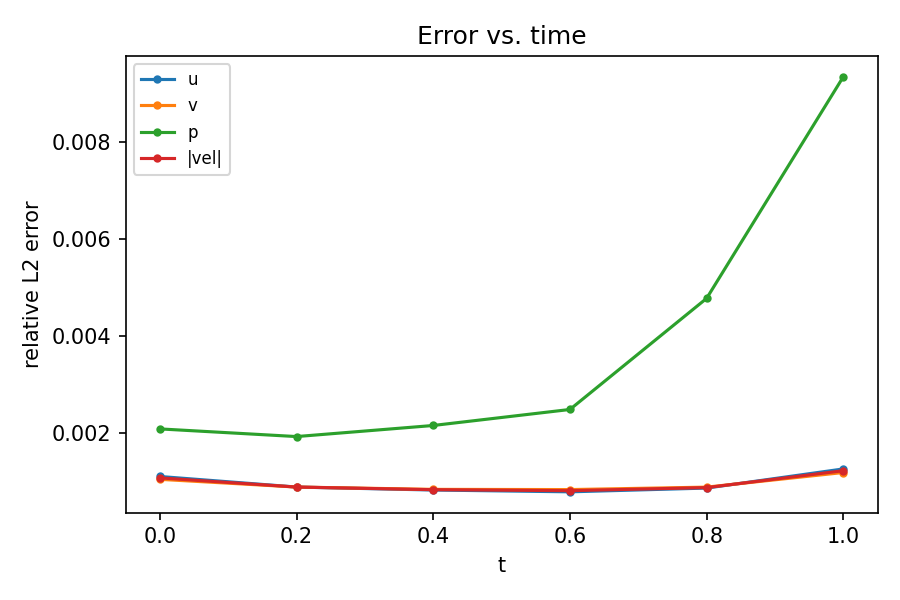}
    \caption{Relative-$L_2$ error vs.\ time.}
    \label{fig:tg-time}
  \end{subfigure}\hfill
  \begin{subfigure}{0.48\textwidth}
    \includegraphics[width=\textwidth]{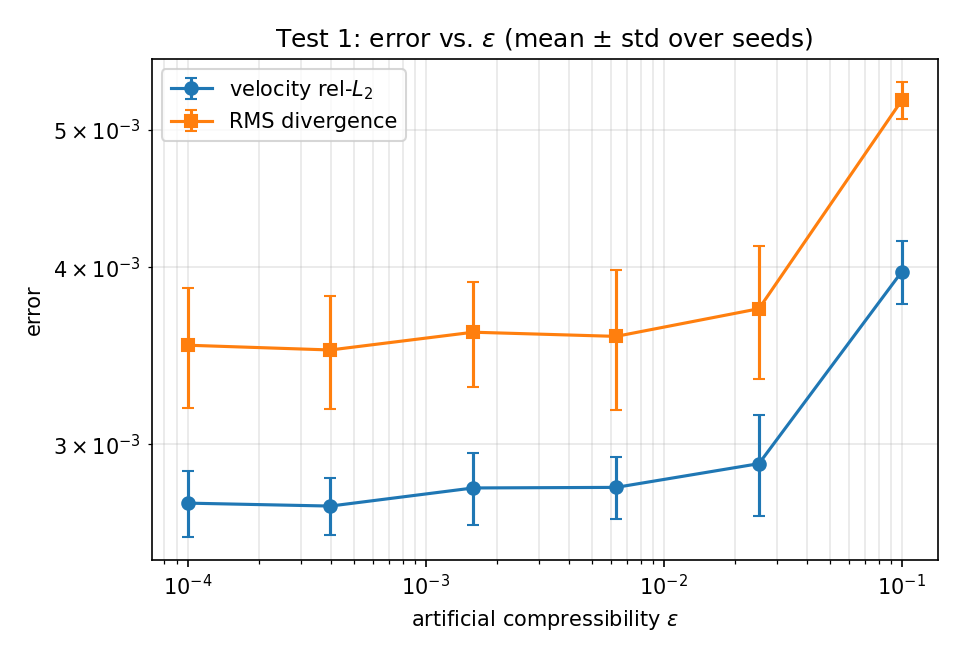}
    \caption{Error and divergence vs.\ $\eps$.}
    \label{fig:tg-eps}
  \end{subfigure}
  \caption{Taylor--Green diagnostics. (a) Per-field error over the simulation
    window. (b) The accuracy--stiffness trade-off in the artificial-compressibility
    parameter $\eps$.}
  \label{fig:tg-diag}
\end{figure}

The AC-PINN reproduces the decaying vortex to a mean relative-$L_2$ velocity error
of $9.5\times10^{-4}$ over $t\in[0,1]$ (final-time error $1.2\times10^{-3}$), with
mean per-field errors $9.5\times10^{-4}$ ($u$) and $9.4\times10^{-4}$ ($v$) for
velocity and $3.8\times10^{-3}$ for pressure; the RMS velocity divergence averages
$1.5\times10^{-3}$. Errors are largest at the temporal endpoints---initialisation
and the final extrapolation horizon---and dip mid-window (Table~\ref{tab:time}).
Figure~\ref{fig:tg-fields} compares the predicted and exact fields at $t=0.6$ and
the pointwise error; Figure~\ref{fig:tg-time} shows the error over time.

\begin{table}[t]
  \centering
  \caption{Taylor--Green relative-$L_2$ error per field and RMS divergence over
    the simulation window.}
  \label{tab:time}
  \sisetup{table-format=1.2e-1}
  \begin{tabular}{S[table-format=1.1] S S S S}
    \toprule
    {$t$} & {$u$} & {$v$} & {$p$} & {RMS div.} \\
    \midrule
    0.0 & 1.10e-3 & 1.04e-3 & 2.09e-3 & 1.84e-3 \\
    0.2 & 8.85e-4 & 8.85e-4 & 1.93e-3 & 1.41e-3 \\
    0.4 & 8.23e-4 & 8.37e-4 & 2.16e-3 & 1.26e-3 \\
    0.6 & 7.86e-4 & 8.32e-4 & 2.49e-3 & 1.25e-3 \\
    0.8 & 8.64e-4 & 8.83e-4 & 4.78e-3 & 1.41e-3 \\
    1.0 & 1.26e-3 & 1.19e-3 & 9.36e-3 & 1.92e-3 \\
    \bottomrule
  \end{tabular}
\end{table}

\subsection{The \texorpdfstring{$\eps$}{eps} trade-off}
\begin{sloppypar}
To isolate the role of $\eps$ we train independent fixed-$\eps$ models (identical
architecture, $4000$ Adam epochs plus a $200$-step L-BFGS polish each, no
continuation) at six values spanning five decades, and repeat each over three
random seeds. Reporting the seed mean $\pm$ standard deviation removes the
optimisation noise that would otherwise contaminate the small-$\eps$ tail.
Table~\ref{tab:eps} and Figure~\ref{fig:tg-eps} show the result. The relaxed
continuity equation~\eqref{eq:ac} forces a residual divergence
$\partial_x u + \partial_y v = -\eps\,\partial_t p$, so the divergence error scales
as $\eps\,|\partial_t p|$; for the Taylor--Green vortex the pressure decays slowly
($\propto e^{-4\nu t}$ with $\nu=0.05$), so even the loosest relaxation
$\eps=10^{-1}$ yields a modest RMS divergence of $5.2\times10^{-3}$ and a velocity
error of $4.0\times10^{-3}$. Reducing $\eps$ lowers both monotonically until they
saturate near $\eps\sim6\times10^{-3}$ at a floor
($\sim2.7\times10^{-3}$ velocity, $\sim3.5\times10^{-3}$ divergence) set by the
network and optimiser rather than by $\eps$. Crucially, once training is
well-converged the small-$\eps$ regime shows \emph{no} stiffness-induced
degradation down to $\eps=10^{-4}$: fixed-$\eps$ AC-PINN training is robust across
the entire range. For reference, the $\eps$-continuation run of
Section~\ref{sec:test1} (four times the epoch budget, with adaptive weighting)
reaches $\eps=10^{-3}$ at a mean velocity error of $9.5\times10^{-4}$, a further
factor of $\sim3$ below the equal-$\eps$ fixed-budget model.
\end{sloppypar}

\begin{table}[t]
  \centering
  \caption{Effect of the artificial-compressibility parameter $\eps$ on the
    Taylor--Green vortex: velocity relative-$L_2$ error and RMS divergence for
    fixed-$\eps$ training, reported as mean $\pm$ standard deviation over three
    random seeds ($4000$ Adam $+\,200$ L-BFGS steps each). The last row is the
    $\eps$-continuation result of Section~\ref{sec:test1}, shown for comparison.}
  \label{tab:eps}
  \begin{tabular}{l c c}
    \toprule
    $\eps$ & velocity $\relL$ & RMS divergence \\
    \midrule
    $1.0\times10^{-1}$ & $3.97\times10^{-3} \pm 2.0\times10^{-4}$ & $5.25\times10^{-3} \pm 1.6\times10^{-4}$ \\
    $2.5\times10^{-2}$ & $2.91\times10^{-3} \pm 2.4\times10^{-4}$ & $3.74\times10^{-3} \pm 4.0\times10^{-4}$ \\
    $6.3\times10^{-3}$ & $2.80\times10^{-3} \pm 1.4\times10^{-4}$ & $3.58\times10^{-3} \pm 4.0\times10^{-4}$ \\
    $1.6\times10^{-3}$ & $2.80\times10^{-3} \pm 1.6\times10^{-4}$ & $3.60\times10^{-3} \pm 3.1\times10^{-4}$ \\
    $4.0\times10^{-4}$ & $2.71\times10^{-3} \pm 1.3\times10^{-4}$ & $3.50\times10^{-3} \pm 3.2\times10^{-4}$ \\
    $1.0\times10^{-4}$ & $2.73\times10^{-3} \pm 1.5\times10^{-4}$ & $3.52\times10^{-3} \pm 3.4\times10^{-4}$ \\
    \midrule
    $10^{-3}$ (continuation) & $9.49\times10^{-4}$ & $1.52\times10^{-3}$ \\
    \bottomrule
  \end{tabular}
\end{table}

\subsection{Which fixes were necessary}
The honest conclusion for this smoothest test is that \emph{none} of the advanced
fixes were strictly necessary. Variable scaling (always on) together with a plain
fixed-$\eps$ residual already reaches sub-$0.5\%$ velocity error at any $\eps$
across five decades, without stiffness pathologies. Gradient-norm adaptive
weighting and $\eps$-continuation each help---the full continuation run of
Section~\ref{sec:test1} reaches $9.5\times10^{-4}$, a factor of $\sim3$ better than
an equal-$\eps$ fixed-budget model and with roughly a third the divergence---but
this improvement comes partly from its larger total epoch budget, and the plain
scheme is already adequate. Causal pseudo-time weighting and augmented-Lagrangian
enforcement were neither used nor needed here. We therefore expect the value of the
toolbox to become apparent only on the harder, genuinely separated flows of
Section~\ref{sec:test23}, where a plain AC-PINN is known to struggle; the
Taylor--Green vortex chiefly serves to validate the formulation and to establish
the $\eps$ scaling law quantitatively.

\section{Test 2: Lid-Driven Cavity}
\label{sec:test2}
The unit-square lid-driven cavity at $Re=100$ ($\nu=U L/Re=10^{-2}$, $U=L=1$) is a
classical benchmark with tabulated reference data~\cite{ghia1982}. The top lid
moves with a smoothed horizontal profile $u_{\text{lid}}(x)=\tanh(50x)\tanh(50(1-x))$
that tapers to zero at the top corners---softening the corner singularities that
otherwise dominate the PINN loss---while the other three walls enforce no-slip.
Starting from rest, the unsteady AC-PINN ($5$-layer, width $128$; $6000$ interior,
$3000$ boundary points; $\eps$-continuation as in Test 1) is marched forward in
time until the flow reaches a statistically steady state, at which point the
vertical- and horizontal-centerline velocity profiles are compared against
Ghia~et~al. No exact pressure is available, so only velocity is supervised on the
walls (per-component boundary weights); the pressure gauge is left free.

\begin{figure}[t]
  \centering
  \includegraphics[width=0.95\textwidth]{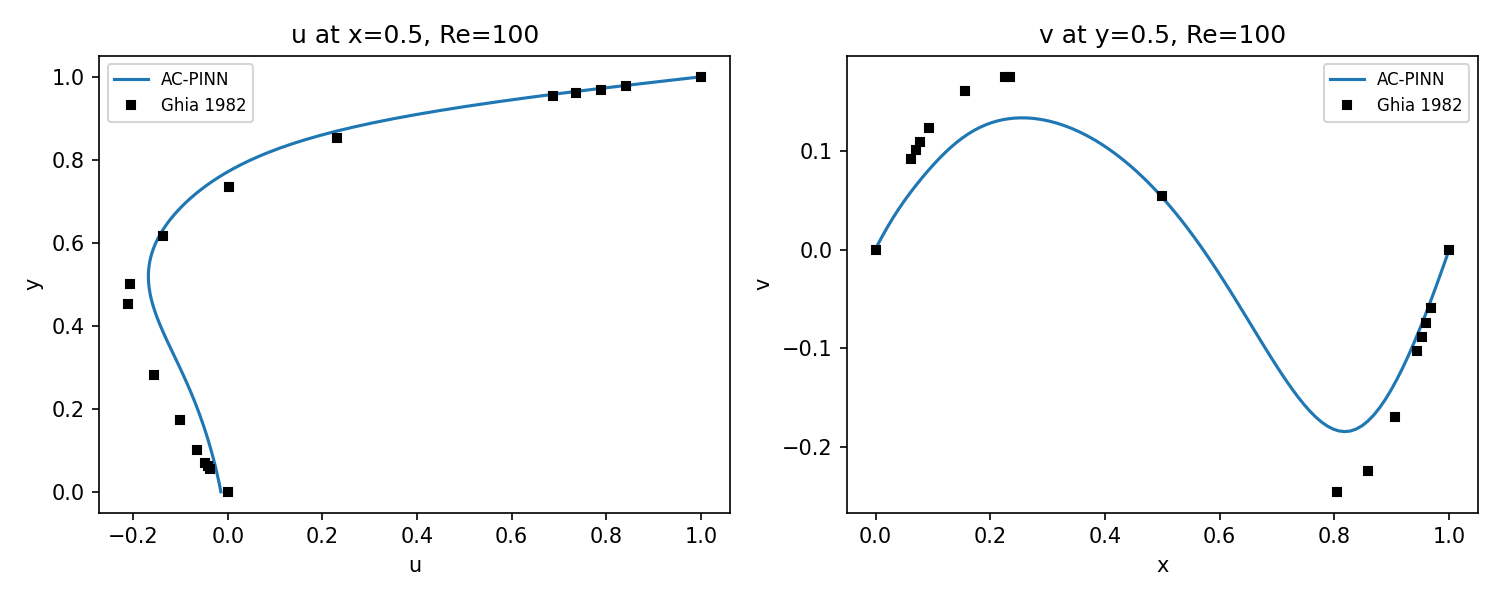}
  \caption{Lid-driven cavity at $Re=100$, statistically steady state ($t=20$):
    AC-PINN centerline velocities against Ghia~et~al.~(1982). Left: $u$ along the
    vertical centerline $x=0.5$. Right: $v$ along the horizontal centerline
    $y=0.5$. Centerline-$u$ relative-$L_2$ error $7.7\%$.}
  \label{fig:cavity}
\end{figure}

\paragraph{Transient development matters.} The primary vortex spins up over
$\mathcal{O}(10)$ convective times. A window of $t\in[0,8]$ leaves the vortex
under-developed---the centerline circulation is systematically $\sim\!30\%$ weak
and the $u$-centerline relative-$L_2$ error against Ghia is $11.9\%$. Extending the
window to $t\in[0,20]$ allows the flow to reach steady state and lowers the error
to $\mathbf{7.7\%}$ (Figure~\ref{fig:cavity}). The recovered $u$-profile tracks the
reference across the full cavity height; the residual discrepancy is concentrated
in the near-wall $v$-extrema, where thin boundary layers appear under-resolved.

\paragraph{Boundary-layer-adaptive sampling does not help here.} A natural remedy
is to concentrate collocation points near the walls. We tried this---both replacing
part of the uniform cloud with a wall-clustered distribution and adding the
wall-clustered points on top of undiminished uniform coverage---and in both cases the
centerline error \emph{worsened} (to $16$--$17\%$). The near-wall deficit is
therefore not a collocation-density problem: the extra near-wall residual terms,
amplified by the gradient-norm weighting, unbalance the objective and degrade the
global circulation that the centerlines measure. We report this as a negative result;
the uniform-sampling $7.7\%$ remains our best cavity accuracy, and closing the gap
likely requires curriculum or hard-constraint treatments of the wall layers rather
than mere point refinement.

\section{Test 3: Cylinder Wake at \texorpdfstring{$Re=100$}{Re=100}}
\label{sec:test23}
Flow past a circular cylinder ($D=1$, centred $4D$ downstream of the inlet in a
$22D\times8D$ channel) produces, at $Re=100$, a periodic von K\'arm\'an vortex
street with Strouhal number $St\approx0.164$--$0.172$. We solve the forward problem
with free-stream inflow and far-field velocity, no-slip on the cylinder, a
pressure-gauge pin at the outflow (mixed velocity/pressure supervision via
per-component boundary weights), and a small antisymmetric transverse perturbation
in the initial condition to break the up--down symmetry. A $7$-layer, width-$160$
network with Fourier-feature inputs ($1.8\times10^{5}$ parameters) is trained with
the full toolbox (scaling, adaptive weighting, $\eps$-continuation, L-BFGS polish).
Sustained shedding is diagnosed from the transverse velocity at a wake probe
($3D$ behind the cylinder): its post-transient RMS and the dominant frequency of
its spectrum.

\begin{figure}[t]
  \centering
  \includegraphics[width=0.95\textwidth]{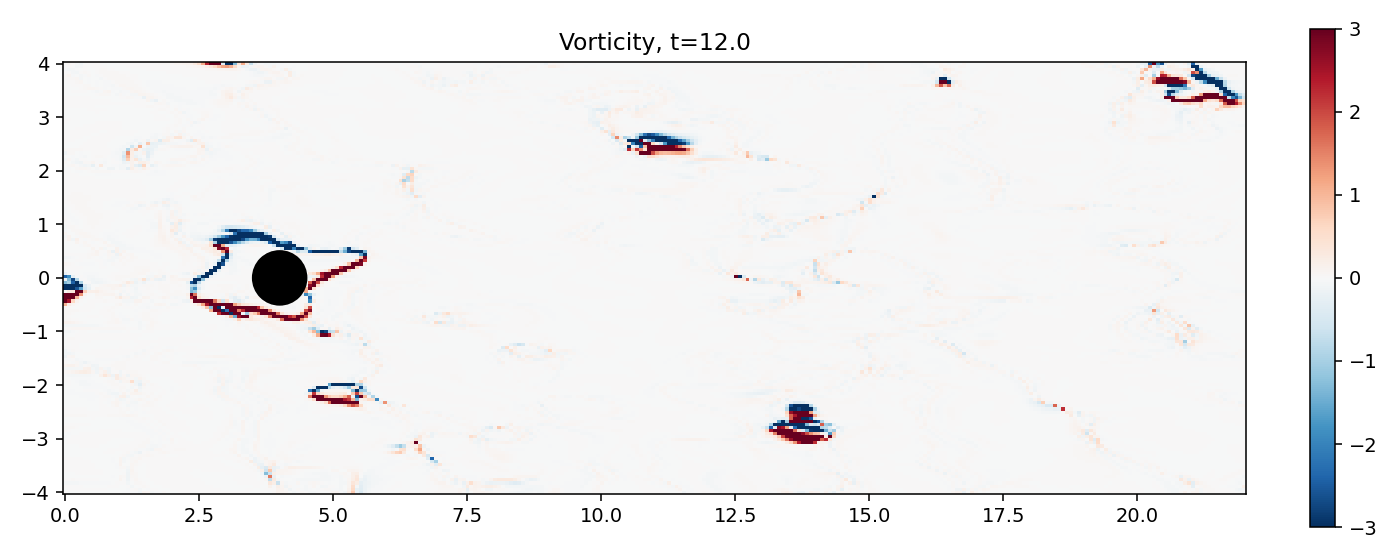}
  \caption{Cylinder wake at $Re=100$, forward AC-PINN, $t=12$. The attached shear
    layers are resolved near the cylinder but the wake does not destabilise into a
    von K\'arm\'an street; downstream vorticity is negligible apart from far-field
    under-resolution artifacts. The plain forward solve collapses to the steady
    symmetric branch.}
  \label{fig:wake}
\end{figure}

\paragraph{The plain forward solve does not shed.} The trained model relaxes to a
\emph{steady, essentially symmetric} attached wake. The probe transverse velocity
has RMS $3\times10^{-3}$ over the second half of the window---two orders of
magnitude below the $\mathcal{O}(0.3)$ oscillation of a genuine vortex street---so
the reported ``Strouhal'' peak ($St\approx0.5$) is spurious noise from an
essentially flat signal, not a shedding frequency. Figure~\ref{fig:wake} shows the
vorticity field: the cylinder boundary layers and the two attached shear layers are
captured, but no coherent vortices are shed downstream, and the far field carries
low-amplitude under-resolution artifacts rather than a periodic wake.

\paragraph{Why, and what it implies.} This is the expected---and, for a methods
paper, the instructive---outcome. The steady symmetric wake is itself a valid
low-residual solution of the governing equations, consistent with reports of the
same collapse for related divergence-free PINN architectures on this
benchmark~\cite{hu2025}, and it sits in a broad basin of
the PINN loss; a weak initial perturbation is smoothed away long before a limit
cycle can organise, because the space-time residual has no mechanism to prefer the
(higher-complexity) unsteady branch. This is precisely the regime the roadmap
reserved for the heavier tools, of which the most reliable is \emph{data
assimilation}~\cite{jin2021nsfnets,franceschini2020}: supplying a few sparse velocity observations so the network is pulled
onto the shedding solution, and additionally inferring a parameter in an inverse
sub-study. We pursue this next.

\subsection{A FEM reference solution}
\label{sec:femref}
To provide assimilation data and a ground truth we solve the same problem with a
standard finite-element method: an incremental pressure-correction (Chorin
projection) scheme with Taylor--Hood-type $P_2$/$P_1$ elements on an unstructured
mesh (CSG rectangle-minus-circle), a smoothly ramped free stream, and a brief
transverse ``kick'' on the cylinder to trigger shedding early. It develops a
sustained von K\'arm\'an street; $100$ velocity snapshots are stored on a regular
grid over a window of $\sim\!6$ shedding periods, together with rings of
boundary-layer probe points just off the cylinder surface (used in
Section~\ref{sec:inverse}).

\paragraph{Getting the Strouhal number right.} A short convergence study
(Table~\ref{tab:st}) is instructive. On the original $22\times8$ domain the
Strouhal number is $St\approx0.18$, above the accepted $Re=100$ value
$0.164$--$0.172$, and---counter-intuitively---\emph{refining the global mesh raises
it} ($0.180\to0.184$) rather than lowering it, while widening the domain leaves it
unchanged. Neither global resolution nor blockage is the culprit: it is the
\emph{boundary-layer} resolution. With only $\sim\!2$ cells across the
$\mathcal{O}(Re^{-1/2})$ layer the separation is mispredicted; locally refining the
mesh to $\sim\!9$ cells across the layer brings $St$ down to $\mathbf{0.176}$ ---
just above the $0.164$--$0.172$ literature band, and by far the closest of our
configurations to it --- with $\overline{C_d}=1.52$ and lift amplitude
$C_l^{\text{amp}}=0.39$. This residual offset is a useful diagnostic in its own
right: the reference frequency used to validate a data-driven wake reconstruction
is only as trustworthy as the boundary-layer resolution of the solver that produced
it, and even a well-resolved reference need not land inside a literature band
whose own scatter reflects differences in domain, blockage, and numerics across
studies. The assimilation study below uses this boundary-layer-resolved reference
on a widened $22\times20$ domain.

\begin{table}[t]
  \centering
  \caption{Strouhal number of the FEM cylinder wake at $Re=100$ under mesh and
    domain refinement. Boundary-layer resolution---not global mesh density or
    domain width---is the controlling factor.}
  \label{tab:st}
  \begin{tabular}{lccc}
    \toprule
    configuration & domain & cells across BL & $St$ \\
    \midrule
    coarse            & $22\times8$  & $\sim1$ & $0.180$ \\
    global-refined    & $22\times8$  & $\sim2$ & $0.184$ \\
    widened domain    & $22\times20$ & $\sim2$ & $0.184$ \\
    BL-resolved       & $22\times20$ & $\sim9$ & $\mathbf{0.176}$ \\
    \midrule
    literature        & ---          & ---     & $0.164$--$0.172$ \\
    \bottomrule
  \end{tabular}
\end{table}

\subsection{Data assimilation recovers the wake}
\label{sec:da}
\begin{figure}[t]
  \centering
  \includegraphics[width=0.95\textwidth]{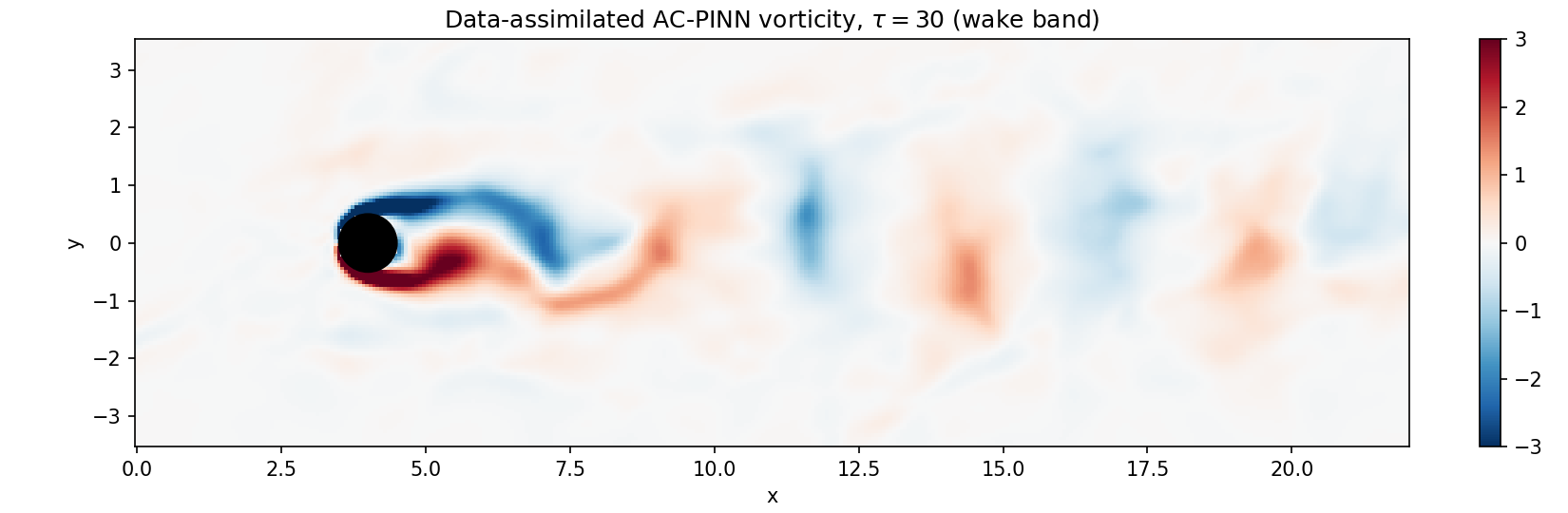}
  \caption{Data-assimilating AC-PINN, wake-band vorticity at $\tau=30$. From sparse
    sensors plus the AC residual, the network recovers a sustained von K\'arm\'an
    vortex street (cf.\ the steady symmetric collapse of the forward solve,
    Figure~\ref{fig:wake}). Wake-band velocity error vs.\ the FEM field: $7\%$;
    shedding frequency $St=0.181$ vs.\ FEM $0.176$.}
  \label{fig:da}
\end{figure}

We place point ``sensors'' at random locations and read the FEM velocity at each of
the $100$ snapshot times, shifting to a PINN time $\tau=t-t_0\in[0,30]$ so the
assimilation window contains fully developed shedding throughout. A data-misfit term
$\|\mathbf{u}_\theta-\mathbf{u}^{\text{obs}}\|^2$ is added to the AC residual and
boundary losses; the network and training are otherwise those of the forward solve.
This single change transforms the result: the AC-PINN reproduces a clean, sustained
vortex street (Figure~\ref{fig:da}), with wake-probe transverse-velocity RMS $0.21$
(versus $3\times10^{-3}$ for the forward solve) and a shedding frequency
$St=0.181$, matching the FEM's $0.176$ to within $3\%$. Both numbers sit on the
same, high side of the accepted literature band ($0.164$--$0.172$; Table~\ref{tab:st}),
so the assimilated PINN inherits the reference's own bias and adds only a small
increment of its own; the $3\%$ figure should therefore be read as a bound on
assimilation fidelity set by the reference solver, not as an independent
validation against the true shedding frequency. Over the active wake band
the velocity reconstruction error against the FEM field is $\mathbf{7\%}$, so a few
hundred sparse sensors plus the physics residual suffice to recover the entire
unsteady flow from a state that a data-free solve could not reach at all --- a
finding consistent with earlier sparse-data reconstructions on related
formulations~\cite{jin2021nsfnets,franceschini2020}.

\subsection{Inverse problem, and where the sensors must go}
\label{sec:inverse}
We now treat the viscosity as an additional unknown (initialised at $\nu=0.02$,
twice the truth) and optimise it jointly with the network. The identifiability of
$\nu$ turns out to depend sharply on \emph{where} the sensors are placed
(Table~\ref{tab:inv}), echoing the sensitivity to observation location documented
for variational data assimilation of unsteady cylinder wakes in the classical CFD
literature~\cite{mons2017}. With sensors only in the wake, $\nu$ descends through the
true value early in training and then drifts on down, settling near
$\nu\approx1.2\times10^{-3}$---almost an order of magnitude too small. The reason is
that the viscous term $\nu\nabla^2\mathbf{u}$ is largest in the boundary layer and
negligible in the wake; wake velocity data leaves the pressure gradient free to
absorb the momentum balance, so the residual is nearly flat in $\nu$ and the
optimiser trades an unphysical viscosity for a marginally better fit.

Moving half of the sensors into the boundary layer---onto rings just off the
cylinder surface, where $\nabla^2\mathbf{u}$ is significant---changes this markedly.
The downward drift is arrested and $\nu$ stabilises at
$\nu\approx6\times10^{-3}$: still low, but a fivefold improvement over the
wake-only estimate and a genuine constraint rather than a runaway. State
reconstruction is unaffected (indeed marginally better). The residual
$\sim\!40\%$ bias is consistent with the remaining pressure/viscous gauge freedom
under sparse data --- comparable in spirit to the identifiability limits reported
for adjoint- and PINN-based viscosity estimators on this benchmark~\cite{zhang2026}
--- and would be expected to shrink further with denser near-wall
sensing or a prior on $\nu$ (e.g.\ along the lines of~\cite{son2023}). The lesson is concrete: for parameter recovery in this
regime, sensor \emph{placement} in the physically informative region matters more
than sensor count, and boundary-layer sensing is what makes $\nu$ identifiable at
all.

\begin{table}[t]
  \centering
  \caption{Inverse recovery of the viscosity ($\nu_{\text{true}}=10^{-2}$,
    initialised at $2\times10^{-2}$) as a function of sensor placement, on the same
    reference and training budget. Boundary-layer sensors make $\nu$ identifiable;
    the state reconstruction is good in both cases.}
  \label{tab:inv}
  \begin{tabular}{lccc}
    \toprule
    sensors & $\nu$ recovered & $\nu$ error & wake field rel-$L_2$ \\
    \midrule
    wake only            & $1.2\times10^{-3}$ & $8\times$ low & $8.6\%$ \\
    wake $+$ boundary layer & $\mathbf{6.0\times10^{-3}}$ & $1.6\times$ low & $7.0\%$ \\
    \bottomrule
  \end{tabular}
\end{table}

\section{Conclusion}
\label{sec:conclusion}
We formulated an artificial-compressibility PINN for the unsteady incompressible
Navier--Stokes equations, relaxing the stiff divergence constraint by a single
parameter $\eps$ so that every governing equation carries a time derivative and the
residual becomes purely local. Across three benchmarks of increasing difficulty a
consistent, honest picture emerges.

On the Taylor--Green vortex the formulation is validated quantitatively: the
residual divergence scales as $\eps\,|\partial_t p|$, both the velocity error and
the divergence decrease monotonically and saturate as $\eps\to0$, and well-converged
fixed-$\eps$ training is stable across five decades of $\eps$ with sub-$0.5\%$
velocity error---no advanced technique is strictly required. On the lid-driven
cavity the unsteady solve reaches the Ghia~et~al. steady state to $7.7\%$
provided the physical time window is long enough for the vortex to develop; here a
smoothed lid and a sufficient time horizon are the decisive ingredients, and the
residual error is localised to near-wall boundary layers. On the cylinder wake the
plain forward solve collapses to the steady symmetric branch and fails to reproduce
shedding, cleanly delimiting the reach of the scaling / weighting / continuation
toolbox; assimilating a few hundred sparse velocity sensors from a
boundary-layer-resolved finite-element reference (Strouhal $0.176$, itself just
above the literature range) then recovers the von K\'arm\'an street to $7\%$ over
the wake and its frequency to $3\%$ of that reference --- underlining that
assimilation fidelity is bounded by reference fidelity. Inferring the viscosity is
harder still and exposes a clean
lesson in conditioning: from wake sensors $\nu$ is unidentifiable and drifts to
nearly an order of magnitude too small, whereas boundary-layer sensors---placed where
the viscous term is significant---make it identifiable and improve the estimate
fivefold, to within a factor of $1.6$.

The practical conclusion is that artificial compressibility makes unsteady
incompressible flow tractable for PINNs on smooth and steady-limit problems with
only lightweight training aids; that self-sustained instabilities are out of reach
for a data-free forward solve but are readily recovered by sparse data assimilation;
and that state and parameter estimation are sharply different in difficulty---the
flow is easy to recover, a physical parameter is not, and where the sensors sit
matters more than how many there are. Two subsidiary findings sharpen the picture:
the reference Strouhal number is set by boundary-layer, not global-mesh or domain,
resolution; and boundary-layer-adaptive collocation, by contrast, does \emph{not}
help the cavity. Immediate next steps are causal pseudo-time weighting and
augmented-Lagrangian enforcement for the forward instability, denser near-wall
sensing or a prior on $\nu$ to close the residual inverse bias, and a drag/lift
validation of the assimilated wake.

\end{document}